\documentclass{ws-procs9x6}
\usepackage{graphicx}

\def\Ladder{
\begin{picture}(160.,50.)
\put(0,0){\line(1,0){150.}}
\put(0,30){\line(1,0){150.}}
\put(155.,0.){\makebox(0.,0.){$S$}}
\put(155.,30.){\makebox(0.,0.){$T$}}
\put(0,0){\line(1,1){30.}}
\put(30,0){\line(0,1){30.}}
\put(60,0){\line(0,1){30.}}
\put(90,0){\line(0,1){30.}}
\put(120,0){\line(0,1){30.}}
\put(30,0){\line(1,1){30.}}
\put(60,0){\line(1,1){30.}}
\put(90,0){\line(1,1){30.}}
\put(120,0){\line(1,1){30.}}
\put(30.,0.){\makebox(0.,0.){$\bullet$}}
\put(30.,30.){\makebox(0.,0.){$\bullet$}}
\put(60.,0.){\makebox(0.,0.){$\bullet$}}
\put(60.,30.){\makebox(0.,0.){$\bullet$}}
\put(90.,0.){\makebox(0.,0.){$\bullet$}}
\put(90.,30.){\makebox(0.,0.){$\bullet$}}
\put(120.,0.){\makebox(0.,0.){$\bullet$}}
\put(120.,30.){\makebox(0.,0.){$\bullet$}}
\put(74.,35.){\makebox(0.,0.){$J_{\parallel}$}}
\put(69.,20.){\makebox(0.,0.){$J_{\rm d}$}}
\put(84.,14.){\makebox(0.,0.){$J_{\perp}$}}
\label{double-chain-2-2}
\end{picture}}

\begin{document}

\title{Theoretical analysis of the double spin chain compound
KC\lowercase{u}C\lowercase{l}$_3$}

\author{M. T. Batchelor, X.-W. Guan and N. Oelkers }

\address{Department of Theoretical Physics,\\ Research School of Physical
Sciences \& Engineering and\\ 
Department of Mathematics,  Mathematical Sciences Institute,\\
Australian National University, Canberra ACT 0200,  Australia}

\maketitle

\abstracts{
We investigate thermal and magnetic properties of the
double spin chain compound KCuCl$_3$ via an exactly solved ladder model
with strong rung interaction. Results from the analysis of the thermodynamic 
Bethe Ansatz equations suggests the critical field values $H_{c1}=22.74\,$T and $H_{c2}=51.34\,$T, 
in good agreement with the experimental observations. The
temperature dependent magnetic properties are directly evaluated
from the exact free energy. Good overall agreement is seen between the theoretical
and experimental susceptibility curves. Our results suggest  that this compound lies 
in the strong dimerized phase with an energy gap $\Delta \approx 35\,$K at zero temperature.  
}
\section{Introduction}

\begin{figure}[t]
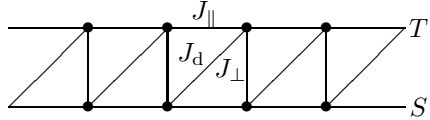

\begin{center}
\Ladder \\
\end{center}
\caption{Schematic picture of the structure of double chain compounds such as KCuCl$_3$.
Here $J_{\perp}$ ($J_{\parallel}$) is the interchain (intrachain) interaction. 
$J_{\rm d}$ is the spin exchange interaction in the diagonal direction.
}
\label{double-chain-2}
\end{figure}

It is believed that the compounds KCuCl$_3$,
TlCuCl$_3$ and NH$_4$CuCl$_3$ exhibit a double spin chain structure,\cite{Tanaka}\cdash\cite{TLCL2}
along the lines of Fig.~\ref{double-chain-2}.  
In the double chain structure, coupling constants $J_{\perp}$ ($J_{\parallel}$) denote 
the interchain (intrachain) spin exchange interactions, with $J_{\rm d}$ a diagonal 
interaction.
However, there appears to be no uniform agreement on the values of these coupling constants 
for the double chain compounds.
In particular, the coupling constants for the compound KCuCl$_3$ are uncertain.
Several theoretical models have been proposed to describe this material, 
including a double chain model with strong antiferromagnetic dimerization,\cite{Nakamura} 
a ladder model with additional diagonal interactions\cite{Tanaka,Tanaka2} 
and a three-dimensional coupled spin-dimer system.\cite{Oosawa}\cdash\cite{Cavadini3} 
None of these models provide an overall fit for all thermal and magnetic properties,   
see, e.g., the review by Dagotto.\cite{Dagotto} 
Measurements of the high field magnetization\cite{Shiramura,Oosawa} and the susceptibility\cite{Tanaka} 
indicate that KCuCl$_3$ exhibits a singlet ground state with an energy gap 
$\Delta \approx 31\,$K at $T=1.7\,$K. 
Nevertheless, it has been difficult to fix all of the coupling parameters
of the model by fitting to only one physical property at a time.
At very low temperatures, $T<5\,$K, the compound KCuCl$_3$ exhibits three-dimensional
magnetic ordering due to complex structural magnetic interaction paths.\cite{Cavadini}\cdash\cite{Cavadini3}

In this communication we investigate the critical fields,
magnetization and susceptibility of the compound KCuCl$_3$ via an integrable ladder model.
The results are used to examine the values of the coupling constants for the double chain structure.
The results for the ladder model with strong rung coupling
are seen to be in good agreement with the experimental results for the energy gap, critical
fields, susceptibility and magnetization.

\section{The integrable ladder model}

It has been shown that integrable (exactly solved) ladder models can be used to
describe real ladder compounds with strong rung interaction.\cite{TBAladder1}\cdash\cite{BGOT}
These integrable ladder models enjoy the nice property
that thermal and magnetic quantities can be obtained exactly
via well developed methods from integrable systems, such as
the Thermodynamic Bethe Ansatz (TBA),\cite{Takahashi}
the Quantum Transfer Matrix (QTM),\cite{QTMrefs} 
$T$-systems\cite{KunibaFusion} and the High Temperature Expansion (HTE) of
Non Linear Integral Equations (NLIE).\cite{ZT2}\cdash\cite{ZT5}

The simplest integrable two-leg spin-$\frac12$ ladder model is constructed from the integrable
$su(4)$ spin chain with singlet rung interaction. 
The Hamiltonian is given by\cite{I-ladder1}
\begin{equation}
{\rm H}=J_{\parallel}{\rm H}_{{\rm leg}}+J_{\perp}\sum_{j=1}^{L}\vec{S}_j\cdot \vec{T}_j-
\mu_{{\rm B}}gH\sum_{j=1}^{L}(S_j^z+T^z_j),
\label{eq:Ham2}
\end{equation}
where
\begin{equation}
{\rm H}_{{\rm leg}}=\sum_{j=1}^{L}\left(\vec{S}_j\cdot \vec{S}_{j+1}+\vec{T}_j
\cdot \vec{T}_{j+1}+4(\vec{S}_j\cdot \vec{S}_{j+1})(\vec{T}_j\cdot \vec{T}_{j+1})\right).
\label{eq:intra} 
\end{equation}
Here $L$ is the number of rungs with $\vec{S}_j=(S_j^x,S_j^y,S_j^z)$ and 
$\vec{T }_j=(T_j^x,T_j^y,T_j^z)$ spin-$\frac{1}{2}$ operators acting on site $j$.
The Bohr magneton is $\mu_{{\rm B}}$ and $g$ is the Land\'{e} factor.
Periodic boundary conditions, $\vec{S}_{L+1} = \vec{S}_1$,
$\vec{T}_{L+1} = \vec{T}_1$, are imposed.

In contrast to the standard Heisenberg ladder model 
the integrable ladder model features an additional biquadratic
spin interaction term in the definition (\ref{eq:intra}) of ${\rm H}_{{\rm leg}}$.
This term causes a shift in the critical value of the rung coupling $J_{\perp}$ 
at which the energy gap closes, and it also causes a rescaling of the parameter 
$J_{\parallel}$ for strong rung coupling. 
In the strong coupling limit $J_{\perp} \gg J_{\parallel}$ the rung
interaction dominates the ground state and low-lying excitations.
The integrable model then lies in the same phase as the standard Heisenberg
ladder, motivating its analysis.

The ground state properties at zero temperature may be obtained from 
the TBA equations.\cite{TBAladder1,mix}\cdash\cite{ying}
Details of the derivation can be found in Ref.~\refcite{BGOT}. 
In the strong coupling limit the integrable spin-$\frac{1}{2}$ ladder model exhibits
three quantum phases: 
a gapped phase in the regime $H<H_{c1}$, 
a fully polarized phase for $H>H_{c2}$ 
and a Luttinger liquid magnetic phase in the regime $H_{c1}<H<H_{c2}$.
The exact values for the critical fields are\cite{TBAladder1} 
$H_{c1}=J_{\perp}-4J_{\parallel}$ and $H_{c2}=J_{\perp}+4J_{\parallel}$.

On the other hand, the temperature dependent free energy has been calculated via
the exact HTE of the NLIE.\cite{HTE1,BGOT}
The free energy of the integrable spin ladder (\ref{eq:Ham2}) is given in the form\cite{HTE1,BGOT}
\begin{equation}
-\frac{1}{T}f(T,H)=\ln Q^{(1)}_1+ \sum_{n=1}^\infty c^{(1)}_{n,0}\left(\frac{J_{\parallel}}{T}\right)^n
\label{eq:freeenergyHTE}
\end{equation}
where $Q^{(1)}$ and the first few coefficients $c^{(1)}_{n,0}$ are given explicitly
in Refs.~\refcite{HTE1,BGOT}.
These terms are functions of the rung coupling $J_{\perp}$, $\mu_BgH$ and the 
temperature.
Most importantly, the exact expression (\ref{eq:freeenergyHTE}) for the free energy can
be used to examine physical properties such as the magnetization, 
susceptibility and magnetic specific heat via the standard thermodynamic relations
\begin{equation}
M= - \left. \frac{\partial f(T,H)}{\partial H} \right\vert_{T}, \quad
\chi =- \left.
\frac{\partial^2 f(T,H)}{\partial H^2} \right\vert_{T}, \quad
C= - T\left. \frac{\partial^2  f(T,H)}{\partial T^2} \right\vert_{H}.
\nonumber 
\end{equation}

\section{The compound KCuCl$_3$}

In this section we examine the low temperature properties of the compound KCuCl$_3$. 
Experimental measurements of the high field magnetization\cite{Shiramura,Oosawa} 
show that magnetic anisotropies are negligible, because the critical fields are almost the same for the
external field in different directions.
However, the susceptibility curves for the external magnetic field along the different
directions are influenced by different $g$-factors.\cite{Tanaka}
In this way magnetic anisotropies may lead to different critical fields for external 
magnetic fields along different directions. 
This can be easily seen from the TBA analysis. 
For instance, if the rung interaction along the $z$-axis is increased,
i.e., by adding an extra term $\Delta_z=\sum_{j=1}^LS^z_jT^z_j$ to the rung
interaction, the critical fields for the magnetic field along the $z$-direction are given by
\begin{eqnarray}
H_{c1}&=&J_{\perp}+\frac{1}{2}\Delta_z-4J_{\parallel},\nonumber\\
H_{c2}&=&J_{\perp}+\frac{1}{2}\Delta_z+4J_{\parallel}.
\end{eqnarray}
{}For the magnetic field along the $x$-direction they are given by
\begin{eqnarray}
H_{c1}&=&\sqrt{(J_{\perp}+\frac{1}{2}\Delta_z-4J_{\parallel})(J_{\perp}-4J_{\parallel})},\nonumber\\
H_{c2}&=&\sqrt{(J_{\perp}+\frac{1}{2}\Delta_z+4J_{\parallel})(J_{\perp}+4J_{\parallel})}.
\end{eqnarray}

The experimental results\cite{Tanaka,Shiramura,Oosawa} indicate that $\Delta_z$ is negligible. 
Analysis of such anisotropic behaviour can be found in Ref.~\refcite{ying}. 
We therefore take the high field magnetization curves for the external field along the
perpendicular and parallel directions to the cleavage plane as evidence that the
double chain ladder model is magnetically isotropic along the chain direction.  
%
%
In the strong coupling case two components of the triplet never contribute to the ground state 
at zero temperature, due to the strong single component contribution along the rungs. 
It has been suggested\cite{FT2} that the triplet excitation can be considered as an analogue
of Bose-Einstein condensation for magnons\cite{BEC1}\cdash\cite{BEC4} for this class of compounds.
The strongly coupled spin ladder with magnon excitations
for strong magnetic fields can be mapped to a one-dimensional $XXZ$-Heisenberg
chain with an effective magnetic field. 
In this case the TBA equations reduce to only one level.
The experimental magnetization curves\cite{Shiramura,Oosawa} suggest an energy gap 
$\Delta \approx 31.1\,$K and the critical field values $H_{c1}\approx 20\,$T and
$H_{c1}\approx 50\,$T at $T=1.3\,$K. 
Fitting the zero temperature TBA critical fields and susceptibility to the 
experimental curves\cite{Tanaka} gives the coupling constants 
$J_{\parallel}=5.5\,$K and $J_{\perp}=57\,$K for the integrable spin ladder model (\ref{eq:Ham2}).

\begin{figure}[t]
\begin{center} 
\vskip 5mm
\includegraphics[width=.90\linewidth]{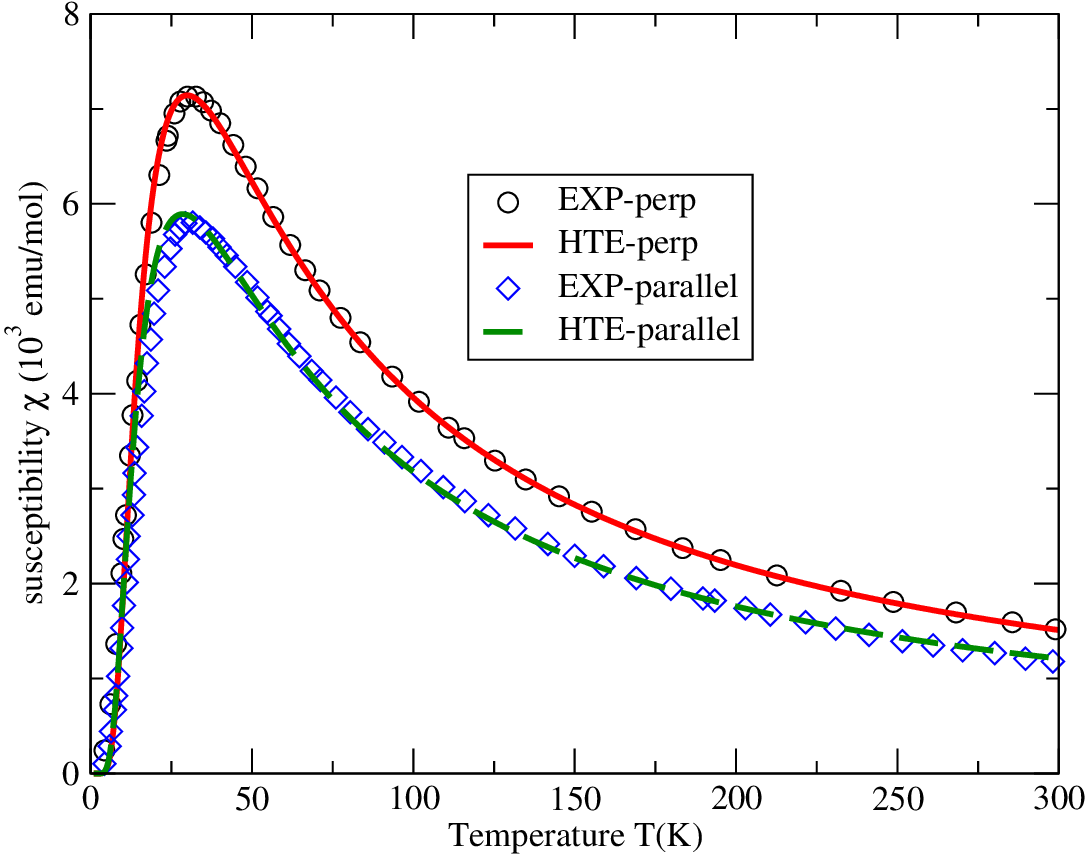}
\caption{
Comparison between theoretical and experimental susceptibility curves 
versus temperature for the compound KCuCl$_3$.
Circles and diamonds denote the experimental data extracted from
Ref.~1 for an external field perpendicular or parallel to the chain
direction.  
The solid and dashed curves are the corresponding
susceptibility curves evaluated directly from the HTE at $H=0\,$T.
Fitting results in the coupling constants $J_{\perp}=57\,$K and
$J_{\parallel}=5.5\,$K with $g=2.29$ (perpendicular), $g=2.05$  (parallel) and
$\mu_B=0.672\,$ K/T.
The conversion constant is $\chi_{{\rm HTE}}\approx 0.40615 \chi
_{{\rm EXP}}$ fixed in Ref.~15. 
}
\label{fig:sus}
\end{center}
\end{figure} 

\subsection{Susceptibility} 

The application of the HTE (\ref{eq:freeenergyHTE}) for the free energy of 
Hamiltonian (\ref{eq:Ham2}) indicates that the 
coupling constants $J_{\parallel}=5.5\,$K and $J_{\perp}=57\,$K
also give excellent fits to the susceptibility.
The temperature dependence of the susceptibility curves is shown in Fig.~\ref{fig:sus}. 
The solid and dashed lines denote the susceptibility
for the external field perpendicular and parallel to the double chain
direction, as derived from the free energy expression (\ref{eq:freeenergyHTE}) 
with up to fifth order HTE.  
Here the Land\'{e} factors $g=2.29$ (perpendicular) and $g=2.05$ (parallel) for the 
external field direction were used. 
A rounded peak at $T=28.5\,$K in the zero magnetic field susceptibility curve
indicates typical antiferromagnetic behaviour.
The overall agreement with the experimental susceptibility curves is excellent.
The susceptibility for the external field parallel to the chain direction has been 
examined via different theoretical models.\cite{Nakamura}
Their conclusion favours a dimerized Heisenberg ladder structure with additional 
diagonal spin interactions, with the suggested coupling constants 
$J_{\parallel}=J_{\rm d}=8.35\,$K and $J_{\perp}=50.1\,$K for the double chain 
structure compound. 
However, their fitting constants result in an energy gap $\Delta \approx 38\,$K, 
which is much larger than the experimental value. 
We conclude that it is not necessary to introduce diagonal spin exchange interaction 
due to the strong dimerization along the rungs. 
The diagonal spin exchange interaction has only a weak effect on the low temperature 
behaviour. 
Moreover, the leg interaction is also suppressed by the relatively strong rung dimerization.

\begin{figure}[t]
\vspace{0.5cm}
\begin{center} 
\includegraphics[width=.90\linewidth]{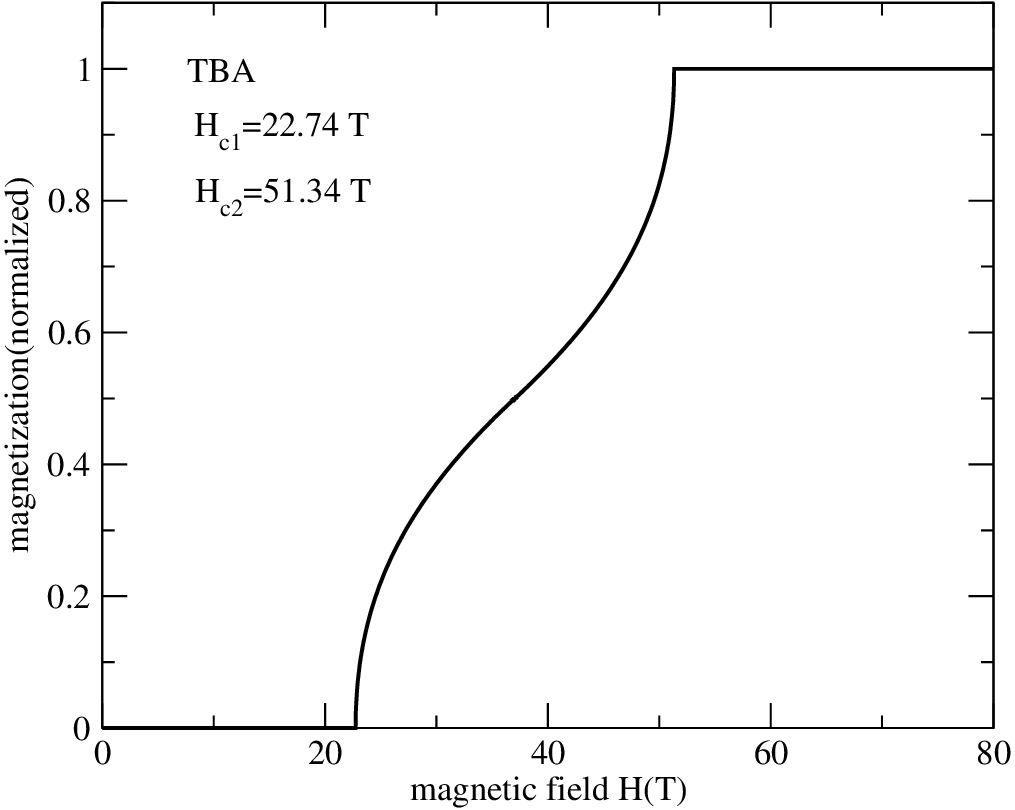}
\caption{
Magnetization versus magnetic field for the compound KCuCl$_3$ with the same constants as in 
Fig.~\ref{fig:sus}.
This curve indicates the nature of  the high field quantum phase diagram. 
The stiffness in the vicinities of the critical fields $H_{c1}$ and $H_{c2}$ is softened by increasing temperature. 
The critical fields predicted by the TBA are in good agreement with the experimental values. 
}
\label{fig:magnetization}
\end{center}
\end{figure} 

\begin{figure}[t]
\vspace{0.5cm}
\begin{center} 
\includegraphics[width=.90\linewidth]{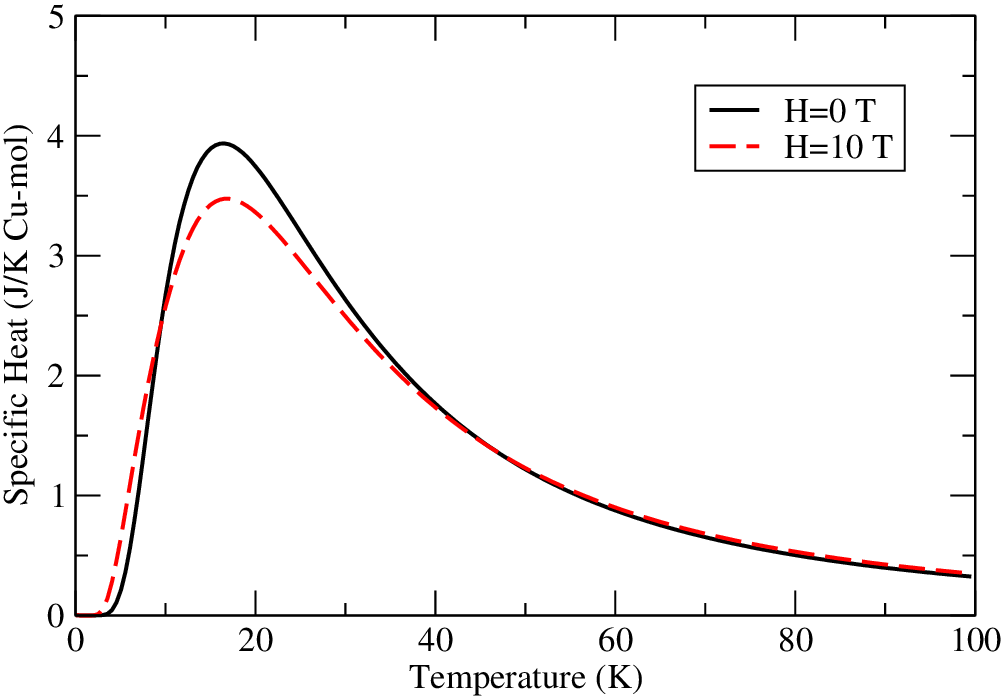}
\caption{
Specific heat versus temperature for different magnetic field strengths for the compound 
KCuCl$_3$ with the same set of coupling constants $J_{\perp}$, $J_{\parallel}$ and $g=2.29$. 
The solid and dashed curves are evaluated from the HTE for $H=0\,$T and $H=10\,$T.
The conversion constant is $C_{{\rm HTE}}\approx 4.515 C_{{\rm EXP}}$.
}
\label{fig:heat}
\end{center}
\end{figure}

\subsection{Magnetization}

The magnetization is a particular interesting quantity to study
as the field dependent magnetization curve leads to the prediction of the low temperature
phase diagram as well as magnetization plateaux.
The high field magnetization curve evaluated from the TBA at zero
temperature is shown in Fig.~\ref{fig:magnetization}. 
By the nature of the high temperature expansion, we are unable to produce these very low 
temperature, $T<5.5\,$K, magnetization curves from the free energy (\ref{eq:freeenergyHTE}) 
for this particular compound.
This highlights the complementary role of the TBA and HTE approaches.
The magnetization curve indicates that the rung singlets form a
nonmagnetic ground state if the magnetic field is less than the
critical field value $H_{c1}=22.74\,$T.
The gap closes at this critical point. 
If the magnetic field is above the critical point, the lower component of the triplet
becomes involved in the ground state. 
The magnetization increases almost linearly with the field towards the
critical point $H_{c2}=51.34\,$T, at which the ground state becomes fully polarized.  
This is in good agreement with the experimental values 
$H_{c1}\approx 20\,$T and $H_{c2}\approx 50\,$T.\cite{Shiramura,Oosawa}

\subsection{Specific Heat} 

Fig.~\ref{fig:heat} shows the specific heat curves obtained from the HTE for the free energy 
at different magnetic field strengths. 
In the absence of a magnetic field the rounded peak indicates short range ordering with a large gap. 
At temperatures less than $T=17\,$K the exponential decay signals an ordered phase. 
The magnetic field is seen to only weakly affect the magnetic specific heat
at low temperatures, mainly because of the strength of the rung singlets.
As yet there appears to be no experimental data for the specific heat.

\section{Conclusions}

We have examined the magnetization, susceptibility and critical fields
of the double chain compound KCuCl$_3$ via the integrable spin ladder model (\ref{eq:Ham2}).
The theoretical results obtained from Thermodynamic Bethe Ansatz and High Temperature Expansion 
calculations are seen to lead to good agreement with the experimental
measurements for these quantities. 
We conclude that this compound exhibits strong rung coupling which leads to dimerized rung spins. 
This is consistent with the experimental analysis.\cite{Tanaka,Tanaka2}
We have also presented the specific heat curves for different magnetic fields.

\section*{Acknowledgments}

M.T.B. and X.W.G. thank Mo-Lin Ge and the Nankai Institute of Mathematics for
their kind hospitality. This work has been supported by the Australian Research Council. 
N.O. has been partially supported by DAAD.

\end{document}